**Brightness Characterization for Starlink Direct-to-Cell Satellites**

Anthony Mallama *[1], Richard E. Cole, Scott Harrington, J. Respler[1]

2024 July 3

[1] IAU - Centre for the Protection of Dark and Quiet
Skies from Satellite Constellation Interference

* Correspondence: anthony.mallama@gmail.com

Abstract

The mean apparent magnitude of Starlink Mini Direct-To-Cell (DTC) satellites is 4.62 while the mean of magnitudes adjusted to a uniform distance of 1000 km is 5.50. DTCs average 4.9 times brighter than other Starlink Mini spacecraft at a common distance.

We cannot currently separate the effects of the DTC antenna itself, the different attitude modes that may be required for DTC operations and to what extent brightness mitigation procedures were in place at the times of our observations. In a best case scenario, where DTC brightness mitigation is as successful as that for other Minis and the DTC antenna does not add significantly to brightness, we estimate that DTCs will be about 2.6 times as bright as the others based upon their lower altitudes.

The DTCs spend a greater fraction of their time in the Earth's shadow than satellites at higher altitudes. That will offset some of their impact on astronomical observing.

1. Introduction

SpaceX launched six Starlink Mini satellites equipped for Direct-To-Cell (DTC) communication on 2024 January 3. Successful testing of that first batch led to the company to request an amendment to their license with the U.S. Federal Communications Commission. That change would allow 7,500 DTC spacecraft to orbit at heights between 340 and 345 km. Those satellites would join 7,500 Starlink Mini satellites at higher altitudes that provide Internet communication. The DTCs are slightly larger than Starlink Internet spacecraft (125 m$^2$ versus 116 m$^2$) due to the DTC antenna.

Satellites are a concern for astronomers because they interfere with observation of the night sky (Barentine et al. 2023 and Mallama and Young 2021). The general impact of Starlink V2 Mini satellites was described by Mallama et al (2024). However, the presence of so many bright DTC satellites at low altitudes would alter those findings. In this paper, we characterize the brightness of DTC spacecraft and re-assess the impact of all Mini satellites.

Section 2 describes how magnitudes were determined for this research and explains the data processing techniques employed to study brightness. Section 3 characterizes the brightness of DTC satellites. Section 4 presents magnitude statistics for the DTC and Internet satellites, and illustrates their distribution in maps of the sky. Section 5 discusses some limitations of this study. Section 6 summarizes our findings.

2. Observations and data processing

The magnitudes analyzed in this study were recorded using electronic and visual methods. The electronic measurements were obtained at the MMT9 robotic observatory (Karpov et al. 2015 and Beskin et al. 2017). The hardware of MMT9 consists of nine 71 mm diameter f/1.2 lenses and 2160 x 2560 sCMOS sensors. MMT9 magnitudes are within 0.1 of the V-band based on information in a private communication from S. Karpov as discussed by Mallama (2021). We collected apparent magnitudes from the MMT9 on-line database along with ranges (distances between the spacecraft and the observer) and phase angles (the arcs measured at the satellite between directions to the Sun and the observer).

The visual observation method, where spacecraft brightness is determined by comparison to nearby reference stars, results in magnitudes that approximate the V-band. The angular proximity between satellites and stellar objects accounts for variations in sky transparency and sky brightness. This method of observing is described in more detail by Mallama (2022).

The distribution of apparent magnitudes for DTC satellites is illustrated in Figure 1. They peak at magnitude 5 but are skewed toward brighter values with a secondary peak at mag 2. The distribution for Internet Mini spacecraft peaks near mag 6 and they are skewed toward fainter mags. The mean apparent mag for DTCs is 4.62 with a standard deviation (SD) of 1.44 and a standard deviation of the mean (SDM) of 0.10. The corresponding values for Internet Minis are 6.36, 0.63 and 0.01. DTC are much more luminous and their brightness dispersion is larger.

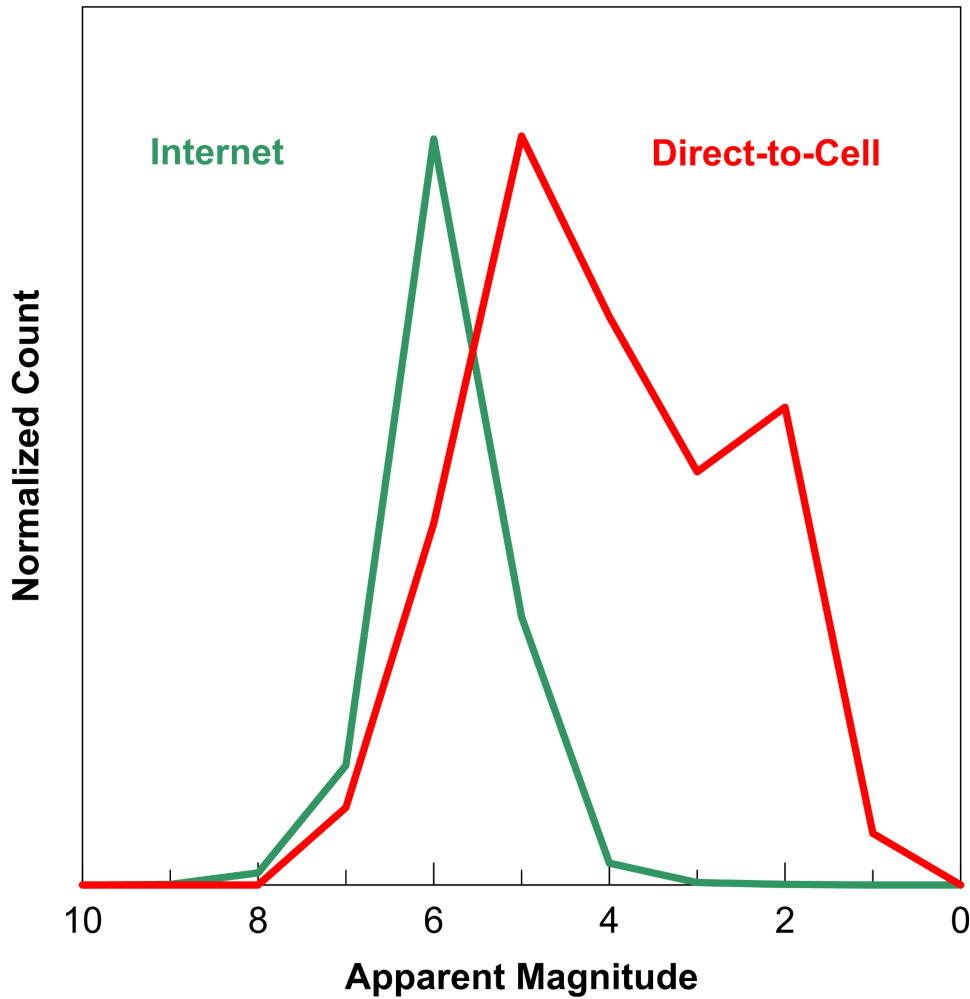

*Figure 1. The distribution of observed apparent magnitudes for Starlink Mini Direct-to-Cell and Internet satellites.*

Data processing includes the computation of ranges and phase angles for each visual observation. Ranges allow magnitudes to be compared at a uniform distance of 1000 km, while phase angles permit the influences of backscattering and forward scattering of sunlight to be assessed. Ranges and phase angles are discussed in the next section.

3. Brightness characterization

When apparent magnitudes for DTCs are adjusted to a uniform distance of 1000 km their mean, SD and SDM are 5.50, 1.38 and 0.10. The corresponding values for Internet spacecraft are 7.22, 0.83 and 0.01. The difference between the mean mags indicates that DTCs are 4.9 times brighter than Internets at the same distance.

The magnitudes and best fitting polynomial phase function for DTC satellites are shown in Figure 2 along with the corresponding polynomial for Internet spacecraft. The function for DTCs is flatter than that for Internets. The DTCs are much brighter than Internets at mid-range phase angle, while they are about the same at large and small angles.

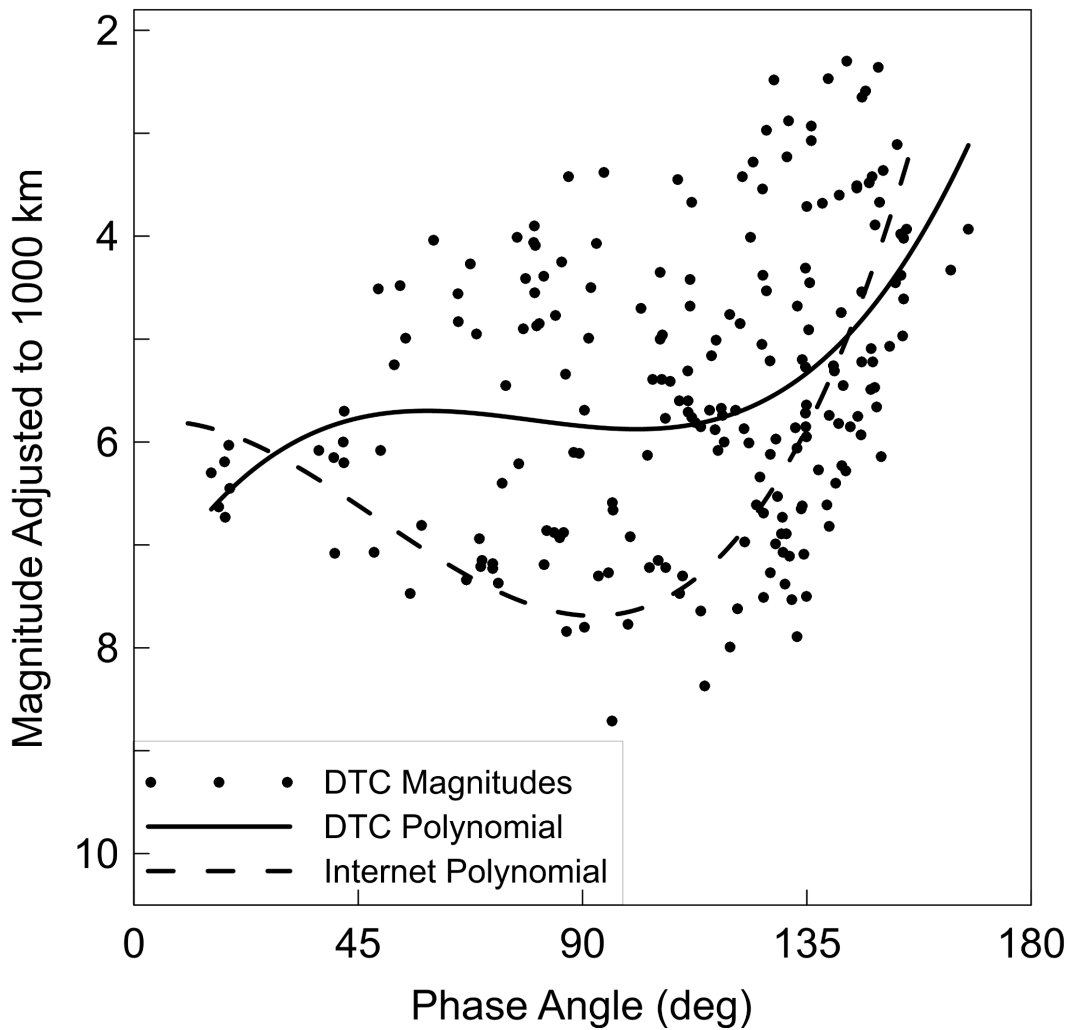

Figure 2. The phase functions for Starlink Direct-to-Cell and Internet satellites.

Table 1. Phase Function Coefficients

| Coefficient | 0 | 1 | 2 | 3 |
|---|---|---|---|---|
| DTC | 7.719 | -0.085395 | 0.001149 | -4.802E-6 |
| Internet | 5.822 | -0.008795 | 0.000848 | -5.784E-6 |

Besides the brightness characteristics described above it is notable that the DTC satellites often appear blue in color. SpaceX informed us that the color is 'not unexpected' but they did not specify which component is blue.

4. Sky maps and magnitude distributions

The phase functions specified in Table 1 were evaluated over a grid of elevation and azimuth to determine apparent magnitudes across the whole sky. Satellite positions were determined by evaluating orbital parameters including altitude and inclination given in the FCC amendment. The dispersions of magnitudes around the best fitting phase functions were applied using a pseudo-random number generator.

The distributions of DTC and Internet satellites are shown separately in Figure 3. In both cases, the observer latitude is 30°, the solar elevation is -18° (the boundary between astronomical twilight and darkness) and the solar declination is 0° (equinox). The DTC spacecraft (left panel) are bright but they are relatively few in number and concentrated toward the horizon. Meanwhile, the Internet satellites (right panel) are spread throughout most of the sky.

Satellites brighter than magnitude 7 cause serious degradation of images for wide field astronomical instruments such as LSST (Tyson et al. 2020), while those brighter than mag 6 are a distraction for visual observers such as amateur astronomers and naturalists. Table 2 indicates that 68 DTC satellites are brighter than mag 7 and 53 are brighter than mag 6. The counts for Internet satellites are 93 and 46, respectively. So, Starlink DTCs may have a greater impact on visual observers than Internets but a lesser impact on wide field instruments.

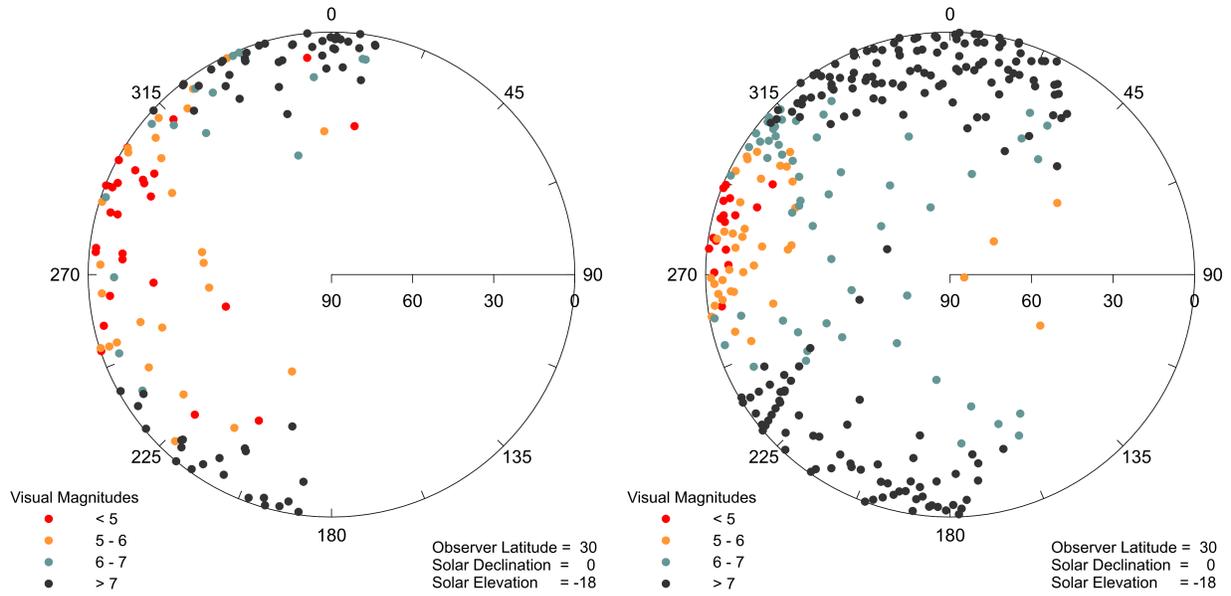

*Figure 3. The distribution and brightness of DTC (left) and Internet (right) Starlink satellites. DTC are largely absent in areas opposite from the solar azimuth (281°) because they are eclipsed by the Earth's shadow. Internet satellites are visible over most of the sky because of their higher altitude.*

Table 2. Simulated Apparent Magnitudes

| Brighter Than | DTC | Internet |
|---|---|---|
| 4 | 7 | 2 |
| 5 | 27 | 18 |
| 6 | 53 | 46 |
| 7 | 68 | 93 |
| 8 | 100 | 148 |
| 9 | 120 | 211 |
| 10 | 129 | 242 |

5. Limitations of this study

While the DTCs are found to be 4.9 times as bright as the Internet Starlinks, we cannot currently separate the effects of the DTC antenna itself, the different attitude modes that may be required for DTC operations and to what extent brightness mitigation procedures were in place at the times of our observations.

In a best case scenario, brightness mitigation for the DTC satellites may be as successful as that for Internet spacecraft and the DTC antenna does not add significantly to luminosity. Then the ratio of brightness may be approximated from the relative altitudes of the DTC and internet spacecraft. Then the DTC would be 2.6 times as bright.

6. Conclusions

The mean apparent magnitude of low altitude Starlink DTC satellites is 4.62 while the mean of magnitudes adjusted to a uniform distance of 1000 km is 5.50. DTCs average 4.9 times brighter than Starlink Internet spacecraft when observed at a common distance.

However, we cannot currently separate the effects of the DTC antenna itself, the different attitude modes that may be required for DTC operations and to what extent brightness mitigation procedures were in place at the times of our observations. In a best case scenario, where DTC brightness mitigation is as successful as that for other Minis and the DTC antenna does not add significantly to brightness, we estimate that DTCs will be about 2.6 times as bright as the others based upon their lower altitudes.

The DTCs spend a greater fraction of their time in the Earth's shadow than satellites at higher altitudes. That will offset some of their impact on astronomical observing.

Final comment

SpaceX recently proposed to orbit 19,440 Starlink internet satellites at a low altitude of 350 km instead of the current 550 km. The distribution in the sky and the apparent magnitudes of these spacecraft are simulated and reported by Mallama (2024).

Acknowledgements

SpaceX provided important information about the status of brightness mitigation for DTC satellites. We thank the staff of the MMT9 robotic observatory for making their data available. The Heavens-Above.com web-site was used to plan observations. Stellarium, Orbitron and Cartes du Ciel were employed to process the resulting data.

References

Barentine, J.C., Venkatesan, A., Heim, J., Lowenthal, J., Kocifa, M. and Bará, S. 2023. Aggregate effects of proliferating low-Earth-orbit objects and implications for astronomical data lost in the noise. Nature Astronomy, **7**, 252-258. https://www.nature.com/articles/s41550-023-01904-2.

Beskin, G.M., Karpov, S.V., Biryukov, A.V., Bondar, S.F., Ivanov, E.A., Katkova, E.V., Orekhova, N.V., Perkov, A.V. and Sasyuk, V.V. 2017. Wide-field optical monitoring with Mini-MegaTORTORA (MMT-9) multichannel high temporal resolution telescope. Astrophysical Bulletin. 72, 81-92. https://ui.adsabs.harvard.edu/abs/2017AstBu..72...81B/abstract.

Karpov, S., Katkova, E., Beskin, G., Biryukov, A., Bondar, S., Davydov, E., Perkov, A. and Sasyuk, V. 2015. Massive photometry of low-altitude artificial satellites on minimegaTORTORA. Fourth Workshop on Robotic Autonomous Observatories. RevMexAA. http://www.astroscu.unam.mx/rmaa/RMxAC..48/PD F/RMxAC..48_part-7.3.pdf.

Mallama, A., 2021. Starlink satellite brightness -- characterized from 100,000 visible light magnitudes. https://arxiv.org/abs/2111.09735.

Mallama, A. and Young, M. 2021. The satellite saga continues. Sky and Telescope, **141**, June, p. 16.

Mallama, A. 2022. The method of visual satellite photometry. https://arxiv.org/abs/2208.07834,

Mallama, A., Cole, R.E., Respler, J., Bassa, C., Harrington, S., and Worley, A. 2024. Starlink Mini satellite brightness distributions across the sky. https://arxiv.org/abs/2401.01546

Mallama, A. 2024. Predicted brightness of Starlink internet satellites at 350 km. https://arxiv.org/abs/2406.16589

Tyson, J.A., Ivezić, Ž., Bradshaw, A., Rawls, M.L., Xin, B., Yoachim, P., Parejko, J., Greene, J., Sholl, M., Abbott, T.M.C., and Polin, D. (2020). Mitigation of LEO satellite brightness and trail effects on the Rubin Observatory LSST. Astron. J. 160, 226 and https://arxiv.org/abs/2006.12417.
7